\documentclass{aa}
\usepackage{graphics}

\begin{document}

   \title{Variability and spectral classification of LMC giants:\\ results from DENIS and EROS
	\thanks {Based on observations collected at the European Southern 
            Observatory}
        \footnote{Tables A.2 and A.3 are only available in electronic form at the CDS via anonymous ftp to cdsarc.u-strasbg.fr (130.79.128.5) or via http://cdsweb.u-strasbg.fr/cgi-bin/qcat?J/A+A/}}
   
   \author{M.-R.L. Cioni\inst{1}
           \and	
           J.-B. Marquette\inst{2}
	   \and
	   C. Loup\inst{2}
           \and
	   M. Azzopardi\inst{5}
	   \and
           H.J. Habing\inst{1}\\
	   \and
	   T. Lasserre\inst{4,3}
	   \and
	   E. Lesquoy\inst{4,2}
          }
 
   \offprints{mrcioni@strw.leidenuniv.nl}

   \institute{Sterrewacht Leiden, Postbus 9513,
              2300 RA Leiden, The Netherlands
         \and
              Institute d'Astrophysique de Paris, CNRS, 
             98 bis Bd. Arago, F--75014 Paris, France
	 \and
	      Max-Planck-Institut f\"{u}r Kernphysik, Postfach 10 39 80,
	      D-69029 Heidelberg, Germany
	 \and
	      Departement d'Astrophysique, physique des Particules, 
	      physique Nucleaire et d'Instrumentation associee (DAPNIA),
	      Service de Physique des Particules (SPP), CEN Saclay,
	      91191 Gif-s/Yvette Cedex, France
	 \and
	      Observatoire de Marseille, LAM, 
              2 place Le Verrier, 13248 Marseille Cedex 4, France 
             }

   \date{Received April 3, 2001/ Accepted April 23, 2001}

   \titlerunning{Variability and spectra of LMC giants}
  
   \authorrunning{Maria-Rosa L. Cioni et al.}

   \abstract{ We  present the first  cross--identifications of sources
        in the  near--infrared DENIS survey and  in the micro--lensing
        EROS survey  in a field  of about $0.5$ square  degrees around
        the  optical center (OC)  of the  Large Magellanic  Cloud.  We
        analyze  the photometric history  of these  stars in  the EROS
        data base and obtain  light--curves for about $800$ variables.
        Most  of  the  stars  are  long period  variables  (Miras  and
        Semi--Regulars),  a few  Cepheids  are also  present. We  also
        present  new  spectroscopic  data  on $126$  asymptotic  giant
        branch stars in  the OC field, $30$ previously  known and $96$
        newly discovered by the  DENIS survey. The visible spectra are
        used  to assign  a  carbon-- (C--)  or oxygen--rich  (O--rich)
        nature to the  observed stars on the basis  of the presence of
        molecular bands  of TiO, VO,  CN, C$_2$. For the  remaining of
        the  stars we used  the ($J-K_S$)  color to  determine whether
        they  are O--rich or  C--rich.  Plotting  $Log(period)$ versus
        $K_S$   we  find   three   very  distinct   period--luminosity
        relations,  mainly  populated   by  Semi--Regular  of  type  a
        ($SRa$),  b ($SRb$)  and Mira  variables.   Carbon--rich stars
        occupy mostly the upper part  of these relations. We find that
        $65\%$  of   the  thermal  pulsing   asymptotic  giant  branch
        population are long period variables (LPVs).
         }

\maketitle

\section{Introduction}
 Stars on the  upper asymptotic giant branch (AGB)  are often found to
be pulsating with large amplitudes and long periods and are classified
as:   Miras,   Semi--Regulars   (SRs)   and  Irregulars.   A   further
characterization in sub--classes is problematic.  A major discovery by
Glass  and Lloyd  Evans (\cite{gle})  is the  linear  relation between
$Log(period)$ ($Log(P)$) and magnitude.  Later it was proven that
O--rich and C--rich AGB stars follow  the same relation in the $K$ mag
(Feast  \cite{feas})  and   in  $M_{bol}$  (Groenewegen  \&  Whitelock
\cite{wg}).   The  $Log(P)$  versus  magnitude relation  is  important
because  it  can  be  used  to  measure the  distance  of  the  parent
population of these stars independently  of other methods.  It is also
important because stellar pulsation modes  are a probe of the internal
structure of the stars and allow quantitative estimates of the stellar
radius,   temperature,  density   and  mass.    Two  years   ago  Wood
(\cite{wood}, \cite{woo}) revolutionized  this picture by showing four
relations for LPVs  in the same diagram. Wood  used data obtained from
the MACHO (Alcock et al.~\cite{alc}) project on the LMC. Very recently
the  combination of MACHO  and ISO  data in  Baade's window  (Alard et
al.~\cite{alal})   in   a   $Log(P)$--magnitude   diagram   showed   a
period--luminosity  relation for  Miras, but  no relation  between the
period  and the  magnitude of  a large  number of  SRs.   An important
aspect of  AGB stars concerns  their atomic composition.   The ``third
dredge--up'',  predicted to  occur in  the thermal--pulsing  AGBs (or:
TP--AGB   phase),  changes  the   surface  chemical   composition,  in
particular the  carbon surface abundance. The  ``dredge--up'' may turn
the star from O--rich  into C--rich (Iben \& Renzini~\cite{iben}).  At
high luminosities (i.e.~$M\approx 5 M_{\sun}$) the formation of carbon
stars  is  prevented  by  the  hot bottom  burning  process  (Iben  \&
Renzini~\cite{iben}). Because  the atmospheres of  O--rich and C--rich
AGB stars are  dominated by different molecular bands,  the stars with
low  mass  loss rate  (such  as those  discovered  by  Blanco et  al.,
\cite{bmb})  occupy   statistically  two  different   regions  in  the
color--color diagram  ($I-J$, $J-K_S$). O--rich AGB  stars have almost
all the same ($J-K_S$), whereas $I-J$ increases with spectral subtype;
TiO molecular  bands strongly affect the  $I$ band -- Loup  et al., in
preparation); C--rich  AGB stars  have a redder  ($J-K_S$) due  to the
effect of the CN and C$_2$  absorption molecular bands in the $J$ band
(Cohen  et al.~\cite{cohen})  and the  $I-J$  ranges from  $1$ to  $2$
magnitudes.   O--rich stars  with high  mass--loss rates,  OH/IR stars
will also have large values of ($J-K_S$).

In this study we inspected  about $800$ EROS light--curves in a region
containing  the optical  center field  (OC) in  the Bar  of  the Large
Magellanic Cloud (LMC); it is one of the fields searched for AGB stars
in the late 1970s by  Blanco et al. (\cite{bmb}). Combining these data
with the DENIS data  we derive three period--luminosity relations.  We
discuss the properties of the stars in the $Log(P)$--magnitude diagram
and  in  the  color--magnitude   diagram  ($J-K_S$,  $K_S$)  with  the
additional information  whether the star  is C--rich or  O--rich. Wood
(\cite{wood}, \cite{woo})  found five  linear relations, we  find only
three because of lower sensitivity in our survey data.

Sec.~$2$   describes   the   DENIS    data,   the   EROS   data,   the
cross-identification between these two data sets and the spectroscopic
observations.  Sec.~$3$ describes  the classification criteria applied
to determine  the variability  and the spectral  type of  the sources.
Sec.~$4$ discusses the period--luminosity relation(s) and the position
of   different   types   of   variables  in   the   ($J-K_S$,   $K_S$)
color--magnitude diagram. Finally  Sec.~$5$ summarizes the contents of
this study.

\section{Data}

\subsection{DENIS photometry}
From  the recently  compiled DENIS  (Deep Near--Infrared  Southern Sky
Survey;  Epchtein et al.~\cite{epal})  point source  catalogue towards
the Magellanic Clouds (DCMC;  Cioni et al.~\cite{cioni}a) we extracted
all the sources overlapping the selected EROS field (Sec.~$2.2$). 
The DENIS  observations have  been made  with the  1m ESO
telescope at La Silla (Chile)  in $1996$ simultaneously in the $I$
($0.8\,  \mu m$),  $J$ ($1.25\,  \mu m$)  and $K_S$  ($2.15\,  \mu m$)
bands;  the   spatial  resolution  is  $1\arcsec$   and  the  limiting
magnitudes are $18$, $16$ and $14$, respectively.  The
DCMC contains only sources detected in at least two of three 
wave  bands.

\subsection{EROS photometry} 
EROS (Exp\'{e}rience  de Recherche d' Objects Sombres)  data come from
observations carried out at the 1m MARLY telescope in La Silla (Chile)
between  August 1996  and May  1999.   A field  of $0.7\degr  (\alpha)
\times  1.4\degr (\delta)$  was  simultaneously observed  in the  EROS
pass--bands ($V_{EROS}$,  $\lambda = 600$ $nm$,  $\Delta\lambda = 200$
$nm$  --  $R_{EROS}$, $\lambda  =  762$  $nm$,  $\Delta\lambda =  200$
$nm$). Each EROS field consists of a mosaic of $8$ $2K\times 2K$ LORAL
CCDs  (one is  out of  use in  the red  camera) with  a pixel  size of
$0.6^{\prime \prime}$.  We used only the central CCDs covering a field
of $0.7\degr (\alpha)\times0.7\degr  (\delta)$ square degrees centered
on $\alpha = 05$:$23$:$24$, $\delta = -69$:$44$:$24$ ($J2000$).
\begin{figure}
\resizebox{\hsize}{!}{\includegraphics{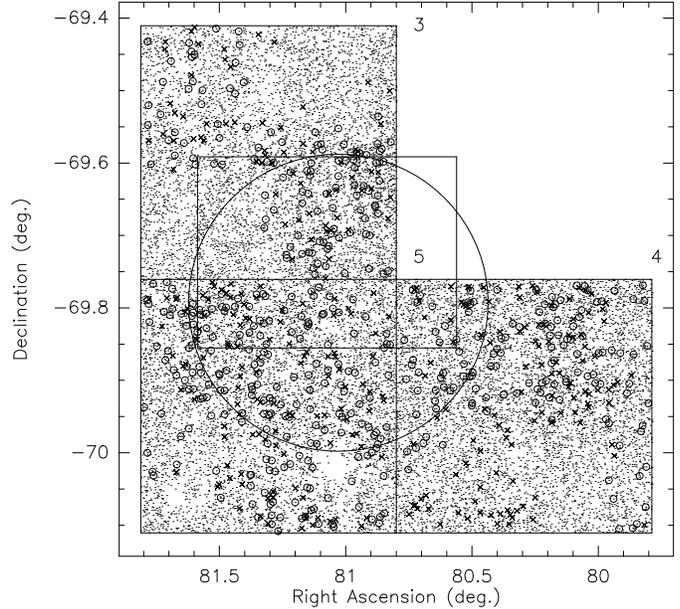}}
\caption{Map of the  analyzed area. 
The big circle limits  the OC field and  the rectangle marks
the area where multi--object spectroscopy was performed.  Small circles
identify  the  classified   variables  and  crosses  identify  
variable stars  without assigned classification.  Small dots represent
all  the  DCMC  sources in  the  field.  The  upper right  empty  area
coincides with the position of the  out of use CCD. Numbers denote the
squared CCDs on the EROS mosaic. The inhomogeneous distribution 
of the variable stars among the CCD quadrants
 is due to the incompleteness of the EROS data.}
\label{mosaic}
\end{figure}
 Flat-fielding,  bias  and  dark  current subtraction)  are  performed
 directly at  the telescope and the subsequent  processing takes place
 in Lyon (France)  at the computing center of the  CNRS.  For each CCD
 quadrant a reference image, that slightly overlaps with the images of
 the adjacent  quadrants, is  composed. Details about  the photometric
 treatment (PEIDA package) are described by Ansari (\cite{ansa}).  The
 OC field is covered by three CCDs and more specifically by a total of
 $8$  quadrants.  Only  a  small  fraction ($10\%$)  of  the OC  field
 overlaps with  the CCD that  is out of use  (Fig.~\ref{mosaic}). Note
 that due to the poor photometry or to technical problems the detected
 variables  are  not  homogeneously  distributed among  the  different
 quadrants.

\subsubsection{Astrometry}
EROS data  have not yet been systematically  calibrated.  We therefore
used  astrometrically the GAIA/Starlink  program and  over-plotted the
DENIS  sources onto  the EROS  images and  applied, in  each  image, a
systematic shift until the EROS  sources covered a plausible subset of
DENIS sources.  We tried to use the USNO astrometric catalogue, but in
the bar  of the LMC the number  of available sources is  not enough to
obtain  a convergent solution.   Therefore we  used all  DENIS sources
detected in three bands and obtained quickly the shift that provided a
very good coincidence between the  two catalogues. We found only a few
double cross--identifications (two EROS  candidates for the same DENIS
source);  in these  cases we  inspected the  corresponding  images and
selected the closer candidate.

\subsubsection{Light--curves}
For each  cross--identified source we extracted  a light--curve.  Next
we searched each  of these light--curve for periods  between $1.5$ and
$1000$ days. The combined extraction  and search program is based on a
method developed by Schwarzenberg--Czerny (\cite{schw}) and calculates
two magnitudes  ($R_{EROS}$ and $V_{EROS}$) and  few other parameters.
One  parameter, ``q''  denotes the  quality of  the  variability; only
sources with $q>25$ are considered to be variable. Sources with $q>50$
have  generally  a  ``good--quality''  light--curve and  sources  with
$25<q<50$ are  clearly variable but  their light curve is  rarely good
enough to  determine any  periodicity; we call  these light  curves of
``acceptable--quality''.  In our sample  $743$ sources are variable of
which $444$ have good--quality and $299$ acceptable quality.  Aliasing
does not affect the extracted light--curves. About $60$ sources of the
good--quality  show  clearly  two  simultaneous periods.   A  separate
detailed analysis,  on all variable sources, where  first the dominant
period is derived and then the  residuals are used to search for other
periods is now being performed (Verschuren et al., in preparation).

\subsection{Optical Spectroscopy}
We  used the  ESO  3.5m New  Technology  Telescope (NTT)  at La  Silla
(Chile)  on the  nights of  November $17$  to $19$,  $1998$  to obtain
low--resolution ($R=270$) spectra  between $\lambda3850-10000$ \AA.  A
sample of $143$  AGB stars in the OC  field in the Bar of  the LMC was
selected from  the DENIS color--magnitude diagram.   For comparison we
added $31$ objects of known spectral type (Blanco et al.~\cite{bmb}).

 We used the Multi--Object--Spectroscopy  mode of the EMMI instrument.
We  observed six  overlapping  fields of  $5^{\prime}\times8^{\prime}$
with  on average  $14$ slits  punched per  mask. The  CCD  frames were
corrected for the bias level introduced by detector electronics (a set
of $9$ frames  per night were observed and  then combined to calculate
the statistical median bias value  for the whole period), for the dark
current (no  dark frames  were observed during  the same period  so we
used  a   value  of  $0.03\,e^-/pix/min$   determined  during  another
observing campaign;  the stability and the almost  negligible value of
the dark  signal is confirmed by  the NTT--team) and  for the relative
pixel response (flat-field --  three frames were observed and mediated
per mask). The sky--subtracted spectra were then wavelength calibrated
and  corrected for  the  atmospheric extinction.   The data  reduction
process was done using the MIDAS package.

Five of the  observed spectra are of poor quality,  but good enough to
 determine  the  C--rich  or   O--rich  nature  of  the  corresponding
 source. $14$ of the observed spectra are useless because the star was
 too faint  or because of  electronic interferences.  In one  case the
 wrong star was observed and in  two cases the slit was in between two
 blended sources.  In the end we  confirmed the spectral  type of $30$
 objects  and  assigned the  spectral  type  to  $96$ objects  of  the
 original sample  (Sec.~4).  Note  that at the  position of  BMB--O 46
 (Blanco et al.~\cite{bmb}) we detected  a star of C type while Blanco
 et al.~assign it a late M type.

\section{Classification of the light--curves}
 With  the   data  provided  by   EROS  we  derive    high--quality
light--curves for sources covering almost the entire Cloud regions and with a
sampling  in time that  was previously  only possible  for very short
period  variables.   The sensitivity  of  the  EROS instrument  allows
amplitudes as small  as $\Delta I = 0.1$ mag to  be measured, a factor
$3$  better  than in  previous  monitoring  campaigns  by i.e.  Hughes
(\cite{hug}).   The  increased  accuracy  of  the  light--curves  also
improves  the  classification  of  the  curve. In  the  GCVS  Kholopov
(\cite{kho})  gives  the  following   criteria:  --  Miras  have  well
pronounced periodicity  in the range  $80-1000$ and a  large amplitude
($\Delta V>2.5$); -- Semi--Regulars  of type a ($SRa$) have persistent
periodicity showing  an amplitude  smaller than Miras;  (amplitude and
light--curve  shape  usually vary  and  the  period  is in  the  range
$35-1200$  days), --  Semi--Regulars  of type  b  ($SRb$) have  poorly
expressed  periodicity (mean cycles  in the  range $20-2300$  days) or
with alternating intervals of periodic and slow irregular changes, and
intervals  of   constant  magnitude;  two  or  more   periods  may  be
simultaneously present.

Frequent  sampling  is  essential  to discover  irregularities  or  to
confirm regularities in the shape  of the light curve. For example the
distinction between $SRa$  and $SRb$ focuses on the  regularity of the
periodicity and  on the  presence of more  than one period,  which can
only be disentangled when the sampling of the light--curve is frequent
enough.  These  better  sampled  light  curves  allow  a  much  better
application of the classification criteria introduced in the GCVS to a
much larger sample of stars.

So far  the detection of small--amplitude variables  in the Magellanic
Clouds  has been  limited.  The  large survey  by  Hughes (\cite{hug})
detected  only  variables with  $\Delta  I>0.5$  mag.  Detailed  light
curves  from  observations  of  variables  in  our  Galaxy  have  been
preferentially determined for large  amplitude variable stars or stars
with infrared excess.

There  are many  more  variable stars  observed  by EROS  that can  be
classified with  refined criteria and  considerably better statistical
completeness. Such studies are outside of the scope of this study.

\begin{figure}
\resizebox{\hsize}{!}{\includegraphics{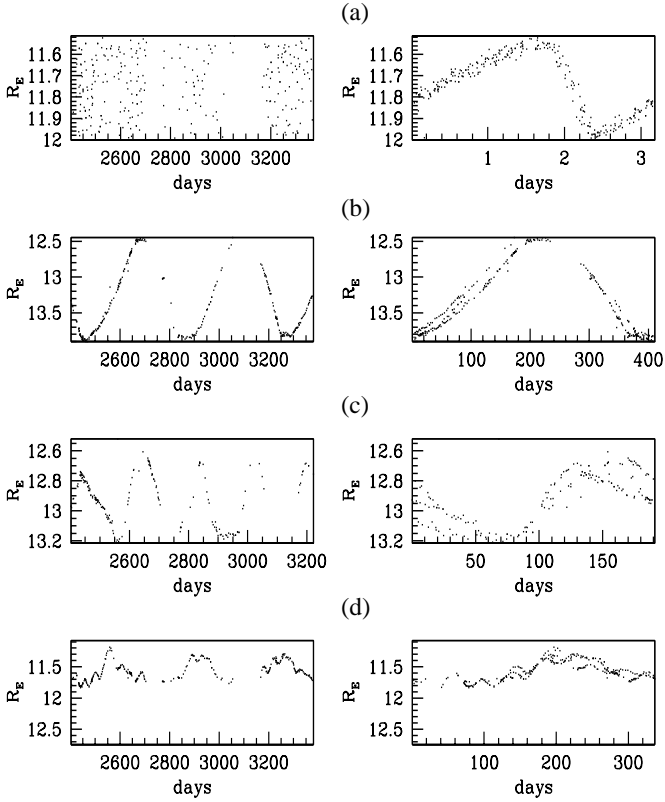}}
\caption{Light--curves  in the  red EROS filter  over the  measured time
range (left) and folded on  the major period (right) for: {\bf (a)} cepheids, 
{\bf (b)} Miras, {\bf (c)} $SRa$s and {\bf (d)} $SRb$s. 
Origin of days for left panels is 
January $1$st, $1990$ ($JD 2,447,892.5$).}
\label{curves}
\end{figure}

\subsection{Cepheid variables}
Cepheids (Fig.~\ref{curves}a)  have periods from $1$ to  $50$ days but
some  larger periods  have also  been  found. We  found $80$  Cepheids
($7.5\%$  of our  total number  of  variable DENIS  stars).  Most  are
detected by DENIS  only in the $I$ and $J$ bands;  their color is blue
reflecting  the  warm temperature  of  the  photosphere ($I-J  \approx
0.7$). In this  paper we concentrate on LPVs and so  we do not further
investigate Cepheids.

\subsection{Long Period Variables}
Mira variables (Fig.~\ref{curves}b) are  AGB stars of spectral type M,
C or S.  They  have a stable period that falls in  the range from $80$
to $1000$  days; their  amplitude exceeds $2.5$  mag in the  $V$ band,
$0.8$ mag  in the $I$ band  (Eggen \cite{egg}) and about  $0.5$ mag in
the near--infrared.  The pulsation  is poorly explained: the pulsation
mode (fundamental,  first or higher overtone) is  not yet established 
 definitively,
nor  is   the  connection   between  mass--loss  and   other  physical
parameters.

We classify $43$ sources as Mira variables adopting the criterion that
$\Delta  R_{EROS}>0.9$, these sources amounting to 
$8\%$ of  the AGB  stars (Fig.~\ref{curves}b).
Their light  curve may show the  following additional characteristics:
-- maxima variable by up to $1$ mag -- small variation of the period from
cycle to  cycle -- presence of  a bump in the  ascending or descending
branch -- variation  of the phase of minimum.   Variables with smaller
amplitude  of variation  have been  classified as  $SR$.   We identify
$25\%$ and $6\%$ of AGB  stars as $SRa$ (Fig.~\ref{curves}c) and $SRb$
(Fig.~\ref{curves}d),  respectively.    $SRb$s  have  more   than  one
pulsation period.  Six $\%$ of  the AGB stars have a light--curve with
too  much  structure   to  identify  them  with  either   of  the  two
sub--classes  of $SR$s.   Among the  acceptable--quality  light curves
there are sources which  can be classified as ``irregular variables'':
clearly variable  but irregular  in amplitude, shape  and size  of the
individual peaks,  therefore without a main period.  Finally there are
also sources with a light--curve of too poor a S/N ratio or a
light--curve which slowly increases or decreases in the whole range of
time. In the latter case the  monitoring period was not long enough to
detect  periodicity; more  data  from  the EROS  or  MACHO (Alcock  et
al.~\cite{alc})  projects  are  certainly  required.  These  last  two
groups of acceptable quality  light--curves, because they indicate that the 
sources are variable but the light--curve is not good enough to assign 
a class of variability, correspond to about $20\%$
of AGB  stars. Table \ref{percent}  summarizes the percentages  of the
different  types of  AGB stars  found in  this work  and  estimated as
described in Sec.~5.
\begin{table}
	\caption{Percentages of AGB stars as discussed in the text.}
	\label{percent}
	\[
	\begin{array}{lr}
	\hline
	\noalign{\smallskip}
	\mathrm{Type}  & \mathrm{\%} \\
        \noalign{\smallskip}
        \hline
        \noalign{\smallskip}
	\mathrm{Mira}    &  8\\
	\mathrm{SRa}     & 25\\
	\mathrm{SRb}     &  6\\
	\mathrm{SR}      &  6\\
	\mathrm{acc.\,quality} & 20\\
	\mathrm{non-variable} & 35\\
	\noalign{\smallskip}
        \hline
        \end{array}
        \]
\end{table}

\section{Spectral classification}
\begin{figure}
\resizebox{\hsize}{!}{\includegraphics{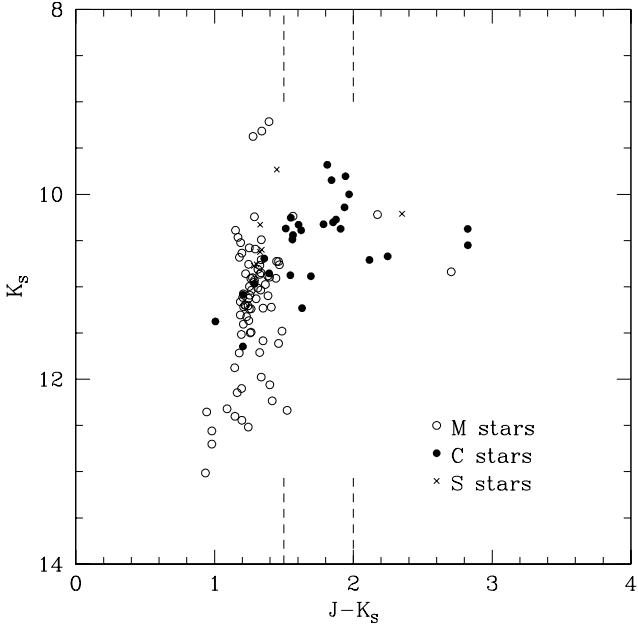}}
\caption{Color--magnitude diagram of the spectroscopically observed DENIS candidates. Filled circles represent C--rich stars, empty circles represent O--rich stars and crosses represent S stars. The vertical dashed lines discriminate between the regions dominated by C--rich, O--rich stars and obscured stars, 
as derived in Sec.~4.}
\label{spec}
\end{figure}

The  spectrum  of  O--rich  AGB  stars is  dominated  in  the  visible
wavelength range by TiO and VO molecular bands. We used these features
to classify  the observed sources  as O--rich.  In particular  the TiO
bands  at $\lambda4761$, $\lambda4954$,  $\lambda5167$, $\lambda5448$,
$\lambda5629$,     $\lambda5661$,     $\lambda5810$,    $\lambda5862$,
$\lambda6159$, $\lambda6187$, $\lambda6596$ (not always very deep) and
$\lambda7055$ \AA, and the VO bands at $\lambda7400$ and $\lambda7900$
\AA  were  used.  Stars  which  have approximately  the  same  surface
abundance of C and O,  ``S--stars'', are characterized by the presence
of molecular absorption bands  of ZrO at $\lambda5551$, $\lambda5718$,
$\lambda5849$,     $\lambda6136$,     $\lambda6154$,    $\lambda6345$,
$\lambda6474$,   $\lambda6495$  and  $\lambda6933$   \AA,  of   YO  at
$\lambda5972$  \AA,  of  TiO  at  $\lambda7055$ \AA  
 (indicating probably a ``MS'' type) ~and  of  LaO  at
$\lambda7404$ and  $\lambda7910$ $\dot{A}$. We  identify C--rich stars
from  the  molecular  absorption  bands of  CN  around  $\lambda6925$,
$\lambda7088$, $\lambda7225$ (triplets), $\lambda5730$, $\lambda5992$,
$\lambda6206$,     $\lambda6332$,     $\lambda6478$,    $\lambda6631$,
$\lambda6656$ and  $\lambda7437$ \AA,  and of C$_2$  at $\lambda4737$,
$\lambda5165$  and $\lambda5636$  \AA.  The  observed sample  of $126$
stars contains $33$ C--rich stars, $7$ S stars and $86$ O--rich stars.
Figure  \ref{spec} shows  the  distribution of  these  sources in  the
color--magnitude diagram ($J-K_S$, $K_s$). Vertical lines indicate the
color  which statistically  discriminate between  O--rich  and C--rich
stars.   TiO and  VO molecular  bands, that  dominate the  spectrum of
O--rich AGB stars, are stronger in the $I$--band than in the $J$-- and
$K_S$--bands;  their ($J-K_S$)  colors  is almost  constant while  the
bands  become  deeper with  increasing  M  type,  and with  increasing
($I-J$). CN and  C$_2$ molecular bands, that dominate  the spectrum of
C--rich AGB stars, are similarly  stronger in the $I$-- and $J$--bands
than in the $K_S$--band.  These stars cover a large range of ($J-K_S$)
colors; it  is not  yet clear  if this is  correlated with  the carbon
spectral subtype.   Objects with $(J-K_S)>2.0$ can  equally be O--rich
or C--rich  stars; relatively massive  and luminous O--rich  AGB stars
with high mass--loss rates have been discovered in the Clouds (Wood et
al.~\cite{woody},   Zijlstra   et   al.~\cite{zijl},   van   Loon   et
al.~\cite{loon}).

For a detailed description  of the features characterizing these stars
and their variation  as a function of the  spectral sub--type we refer
the reader to the Atlas of  Digital Spectra of Cool Stars (Turnshek et
al.~\cite{turn}).

\section{Statistics}
All DCMC sources with few exceptions have an EROS counterpart, because
 EROS is  deep enough ($I\approx20\div21$  mag) to detect  the sources
 revealed at  least in two DENIS  wave bands. Not  all identified DCMC
 sources are  variables. This allows  us to estimate what  fraction of
 the AGB  stars, present  in the DCMC,  are variables within  the EROS
 completeness  limit.  For  statistical  purposes we  focused on  four
 quadrants (equivalent  to one CCD  area) where all DCMC  sources were
 successfully   cross--identified  with   EROS  sources,   areas  that
 therefore  have not  been affected  by technical  problems.   In each
 quadrant  there  are on  average  $1500$  DCMC  sources of  which  we
 identify  $100$  to  be  in  the  AGB  phase,  i.e.~  $K_S<12.0$  and
 $(J-K_S)>1.2$.
 
\begin{figure}
\resizebox{\hsize}{!}{\includegraphics{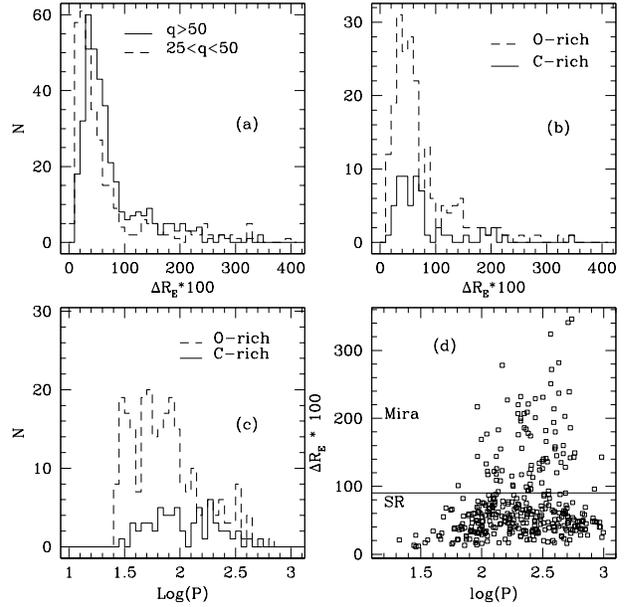}}
\caption{Histogram distribution of the amplitude of {\bf (a)} all the variable sources. For good--quality light--curves both {\bf (b)} the distribution of amplitudes and {\bf (c)} periods are shown as a function of O--rich or C--rich type. {\bf (d)} shows how the amplitudes distributes as a function of $log(P)$, the horizontal line indicates the applied selection criterion to distinguish Miras from $SR$s.}
\label{amp}
\end{figure}

Eight  $\%$  of  the  all  cross--identified  sources  show  signs  of
 variability   in   a   good   ($q>50$)  or   acceptable   ($25<q<50$)
 light--curve.   Among the  AGB stars,  as defined  above,  $65\%$ are
 variable ($q>25$).  Changing the  limits to characterize the location
 of the  AGB stars in  the color--magnitude diagram we  obtain: $50\%$
 for $(J-K_S)>1.0$ ($61\%$ if  we also require $K_S<11.5$), $65\%$ for
 $(J-K_S)>1.6$, $50\%$  for $(J-K_S)>2.4$.  From the  histogram of the
 distribution   of   the   amplitudes   of  the   detected   variables
 (Fig.~\ref{amp}a)  the EROS  data are  complete up  to  about $\Delta
 R_E=0.3$.

\subsection{Amplitude of variability}
Figure \ref{amp}a shows the distribution of the amplitude of the $334$
variable  sources  detected  at  least  in $K_S$.   The  peak  of  the
distribution is around  $\Delta R_E = 0.5$ mag for  the sources with a
good--quality     light--curve,     classified     and    shown     in
Fig.~\ref{per}.  Variable  sources   without  a  well  defined  period
(irregular,  poor  S/N or  monotonic)  peak  at  about $\Delta  R_E  =
0.3$. These  peak positions roughly  indicate the completeness  of the
EROS  data for  the  chosen DCMC  sample.   Fig.~\ref{amp}b shows  the
amplitude  distribution  of  the  good--quality  light--curve  sources
distinguishing between C--rich and O--rich type.  There is no apparent
indication  of  a  difference  in  amplitude between  the  two  types.
Fig.~\ref{amp}c shows  the period distribution  also as a  function of
spectral  type.  O--rich AGB  stars  dominate  at $Log(P)<2.0$,  while
O--rich and C--rich AGB stars are equally abundant at $Log(P)>2.0$. On
average O--rich AGB  stars in this sample tend  to have shorter period
than C--rich  stars; the latter  have an almost  constant distribution
for $1.5<Log(P)<2.5$.  Finally Fig.~\ref{amp}d shows  the distribution
of the amplitudes as a function of $Log(P)$ and the separation between
Miras and Semi--Regulars in our sample.

\section{Discussion}
\begin{figure}
\resizebox{\hsize}{!}{\includegraphics{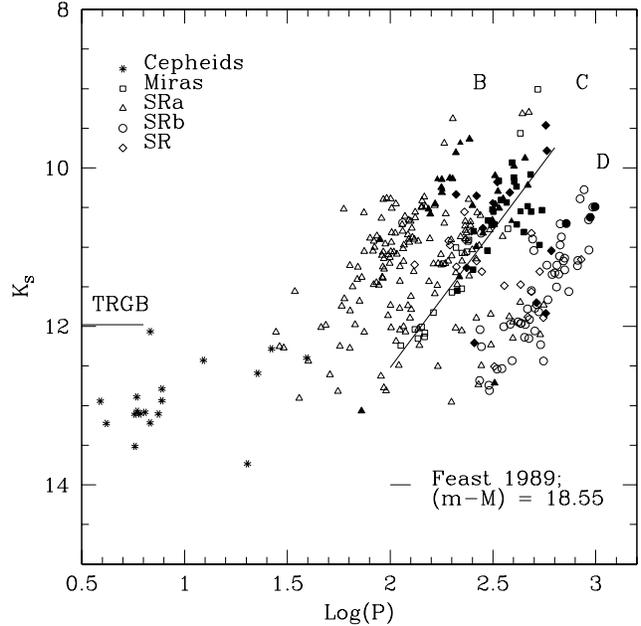}}
\caption{Period--magnitude    diagram   for    the good--quality
 light--curve variable  stars; the period ($P$) is in days.  
Symbols  refer  to  different  types  of
variable sources  and are explained in  the plot. Only the long period 
of $SRb$s is plotted in this figure, see Fig.~\ref{double} for the location 
of the two derived periods. 
Filled symbols are 
C--rich objects ($J-K_S>1.5$) and empty symbols are O--rich objects.
The  position of the
tip  of  the   red  giant  branch  (TRGB)  as   derived  by  Cioni  et
al.~\cite{cionitip}   is   indicated.    The tilted line represents the
period-luminosity relation derived by  Feast (\cite{feas}) for Miras in
the Galaxy  translated in the LMC at $(m-M)=18.55$. 
Capital letters identify 
the sequences as in Wood et al. (\cite{wood}).}
\label{per}
\end{figure}

Fig.~\ref{per} presents the $K_S$  magnitude as a function of $Log(P)$
 for  the sources with  a good--quality  light--curve.  We  ignore the
 cepheids in  the following discussion.  Long  period variables occupy
 the  region of $Log(P)>1.4$.   Three well  defined groups  of objects
 delineate  three sequences  of increasing  magnitude  with increasing
 period labelled  $B$, $C$ and $D$ reflecting  the correspondence with
 Wood et al.~(\cite{wood}): $SRa$s  are mostly distributed in relation
 $B$, Miras and  those $SRa$s that have $\Delta  R_{EROS}>0.7$ are all
 concentrated  in relation  $C$  and $SRb$s  (they  have two  periods)
 occupy relations $B$ and $D$.

The location of Miras is consistent with the LMC Mira sequence of Feast 
(\cite{feas}). The distribution of $SRa$s is also consistent with what
has  been found  in the  Galaxy  (Bedding \&  Zijlstra \cite{bz})  and
previously  in the Magellanic  Clouds (Wood  \& Sebo  \cite{ws}).  The
longer periods of  the $SRb$s give rise to a new  relation so far only
discussed by Wood (\cite{wood}, \cite{woo}), his sequence $D$.

There are few  more objects in Fig.~\ref{per} with  short periods that
correspond to  a sequence found  by Wood and  labelled by him  as $A$.
Both Wood's  sequences $A$  and $B$ are  densely populated  at fainter
magnitudes,  while  our sequences  are  better  populated at  brighter
magnitudes. We attribute this  difference to the limiting magnitude of
DENIS (Cioni et al.~\cite{cioni}) and to the incompleteness of EROS in
detecting small amplitude variables.   The sequence of shorter periods
of $SRb$s, confirmed to be mostly on sequence $B$, coincides also with
the short period  stars found in the Baade's window  by the ISOGAL and
MACHO  collaboration (Alard  et al.  \cite{alal}).   Wood's additional
sequence  $E$ of  eclipsing  binaries is  definitely  below the  DENIS
limiting magnitudes.

 It is important  to emphasize that the $K_S$ magnitude  of DENIS is a
single epoch  measurement and thus the  $Log(P)$--magnitude diagram is
affected  by the scatter  due to  the amplitude  of variation  of each
source; this effect is larger for Miras than for Semi--Regulars.

\begin{figure}
\resizebox{\hsize}{!}{\includegraphics{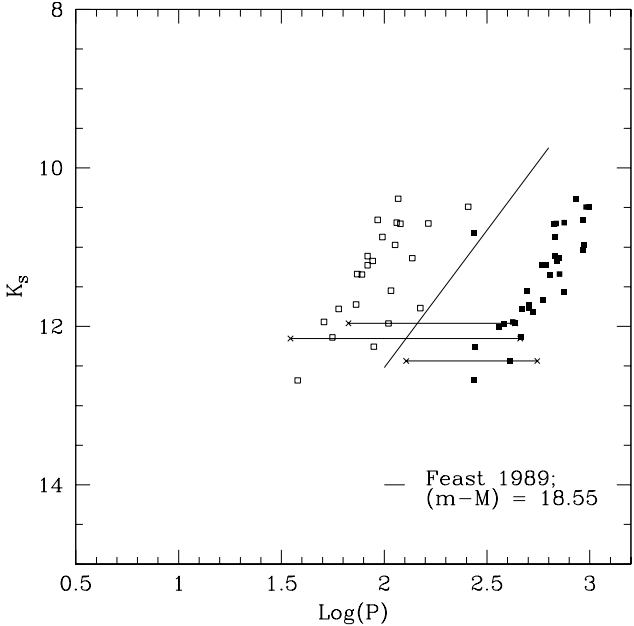}}
\caption{Period--magnitude diagram of $SRb$s. Filled symbols show the long 
period and empty symbols show the short period for the same source. 
Horizontal lines connect possible eclipsing binaries. 
The period ($P$) is in days.}
\label{double}
\end{figure}

\begin{figure}
\resizebox{\hsize}{!}{\includegraphics{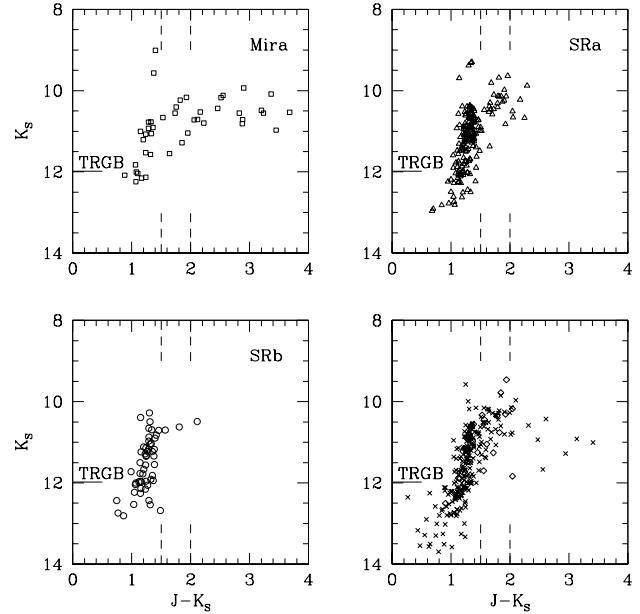}}
\caption{Color--magnitude diagrams ($J-K_S$, $K_S$) of the detected 
variables. Symbols are as in Fig.~\ref{per} and crosses indicate sources of 
acceptable--quality light--curve; Cepheids are excluded from these plots.}
\label{cmd}
\end{figure}

Each  period--luminosity  relation  corresponds, theoretically,  to  a
 region of instability which is  normally associated to a certain mode
 of   pulsation;  the   presence  of   multi--periods   indicates  the
 coexistence  of  two  pulsation  modes.   Fig. 5  of  Wood  and  Sebo
 (\cite{ws})  indicates that the  long period  in $SRb$s  results from
 radial  fundamental mode  pulsation, while  Miras are  first overtone
 pulsators and  other Semi--Regulars  are mostly high  radial overtone
 pulsators; the  predicted gap between fundamental  and first overtone
 sequences is in  very good agreement. This would  also agree with the
 pulsation  mode  derived  by  Whitelock  \&  Feast  (\cite{wf})  from
 measured diameters (Hipparcos data) of Miras in the Milky Way. On the
 other hand if  Miras are fundamental mode pulsators,  as suggested by
 Wood et  al.~(\cite{wood}), then the  long period sequence  of $SRb$s
 requires another  explanation, because theoretically  there exists no
 pulsation mode at a longer period than that of the fundamental mode.

The presence of two simultaneous periods in $SRb$s may be explained by
 the coupling  between dust formation  and star pulsation  (Winters et
 al.~\cite{wint}) because of a  pulsation of longer period produced by
 the dynamical  structure of the circumstellar dust  shells.  As shown
 by Fleischer  et al.~(\cite{flei}) for  large carbon to  oxygen ratio
 ($C/O >  1.8$) the periodic ejection  of a dust  shell occurs exactly
 with the  pulsational period  while for $C/O<1.5$,  as the  amount of
 dust  that can  condense is  smaller, shell  ejection takes  place at
 longer  intervals than one  period (Fleischer  et el.~\cite{flewin}):
 the pulsation corresponds  to the shorter of the  two periods and the
 longer is  due to  obscuration by dust.   It is  however questionable
 whether such  a model  applies to stars  with silicate type  dust and
 lower mass--loss rates ($10^{-6}$ to $10^{-8}$ $M_{\sun} yr^{-1}$) such as
 those  derived  from CO  observations  (Loup  et al.~\cite{lou}) but 
 preliminary  calculations  support that  the model does apply  
(Winters,  private
 communication). If there is  circumstellar dust the stars is expected
 to  have  redder ($J-K_S$)  than  those  observed  by DENIS  in  this
 sample.  On  the  other  hand  if  at  these  colours  the  molecular
 blanketing effect (in the $J$ band due to C$_2$ and CN) dominates the
 dust obscuration effect the blue ($J-K_S$) colour is explained.

Wood et  al. (\cite{wood}, \cite{woo})  suggests that sequence  $D$ is
populated by the longer period of a binary system.  Such a scenario is
also poorly constrained.   We found in our sample  only three possible
eclipsing  binaries  (Fig.~\ref{double}).    Moreover  the  number  of
objects  populating  this  sequence   is  too  high  with  respect  to
theoretical predictions. 
 Whitelock   (\cite{white})   shows,  from
measurements in globular clusters,  that the evolutionary path of long
period  variables  goes  from   $SRa$  to  Mira  following   a line 
of shallower slope, in agreement with the theoretical calculations by 
Vassiliadis and Wood  (\cite{vass}), than the period--magnitude relation.  
 The evolutionary
path thus shows that the stars pass through a variable phase, then the
pulsation goes away and they slowly move into the next sequence.  Gaps
between  the sequences  reflect a  ``phase'' of  some  stability which
could be represented by the $35\%$ of non--variable AGB stars. However
because of  the existing gap between  the starting of  the core Helium
burning (TRGB) and the beginning of the TP--AGB phase (Iben \& Renzini
\cite{iben})  some of  the AGB  stars  in our  sample might  be in  an
earlier evolutionary phase (E--AGB)  or they are TP--AGB stars in between 
thermal pulses that esperience a significant decline of luminosity 
(Wood et al.~\cite{wood}).  Little evidence of pulsation has
been found  in AGB  stars just above  the horizontal branch  (Feast \&
Whitelock \cite{fw}), these E--AGB stars  might be of the same type as
those  AGBs we do  not detect,  within the  completness limits  of the
sample, as variables.  Alard et al.~(\cite{alal}) found that $92\%$ of
the stars in their sample are variable. These stars have been selected
to  have a  detection both  at $[7]\mu$m  and $[15]\mu$m  (ISO bands).
Because of the detection at $[15]\mu$m, that probes the existence of a
circumstellar dust envelope (Omont et al.~\cite{omont}), we think that
Alard et al.~sample is formed  by TP--AGB stars which are 
therefore more evolved
than our sample  of AGB stars selected on  the basis of near--infrared
colours. It is thus not surprising  that we find a lower percentage of
AGB variables. However it is remarkable that in our sample only $50\%$
of  the high  mass--loss stars  ($(J-K_S)>2.4$) are  variable, because
mass--loss and  variability are believed  to be connected.  The result
may be due  to incompleteness of the sample  (Sec.~5.1); note that the
sample size  for stars  with  ($(J-K_S)>2.4$) is  much smaller  because
DENIS does not detect very obscured AGB stars.

 The   color--magnitude    diagrams   ($J-K_S$,   $K_S$)    shown   in
Fig.~\ref{cmd}  characterize the  location of  the different  types of
variable. Miras cover uniformly  the region where O--rich, C--rich and
obscured  stars  are  located  (Fig.~\ref{spec}). Comparing  the  four
diagrams, the  reddest objects  are all of  Mira type. Not  all Miras,
however, have  red colors. This  may indicate a difference  in initial
mass.   The initial  ``position'' (luminosity,  period, color)  of the
stars in  the TP--AGB  phase is given  when they experience  the first
TP. All TP--AGB stars loose mass  at a rate that varies from $10^{-8}$
to $10^{-4}$ $M_{\sun} yr^{-1}$ and will end as a $0.6\div 1.0 M_{\sun}$
white dwarf. Because the time on the AGB of the less massive stars is
much longer than that of the more massive stars, the mass loss rate of
a low mass star can be much lower. As a consequence in the last stages
of the AGB (Mira) low mass  stars do not become much redder because of
mass loss, but the more massive stars will.  The AGB phase is so brief
(Vassiliadis \&  Wood \cite{vass}) that there is  very little increase
in luminosity.  Stars that are  too light will remain  O--rich because
the TPs have too little effect. Stars that are too massive will remain
O--rich because of hot--bottom  burning. Only stars in a well--defined
mass  interval will  become  C--rich.  In the  LMC  this explains  why
C--rich  stars  have  almost  all  the  same  luminosity  (Groenewegen
\cite{gro}).  $SRa$s  are distributed in both branches  of O--rich and
C--rich stars while $SRb$s are mostly concentrated only in the O--rich
branch  (Fig.~\ref{cmd}).   Sources with  good  or acceptable  quality
light--curves,  but no  classification, are  distributed  in different
regions of the diagram and their type can probably be derived from the
diagrams of the classified sources.

$SRb$s  might be  in a  transition  phase between  O--rich $SRa$s  and
Miras.   At the  beginning and  at the  end of  the TP--AGB  phase the
amount of material  available to form dust is high,  in the first case
because the excess of oxygen is  high and in the other one because the
excess  of   carbon  is  high.    We  expect  such  objects   to  have
light--curves like $SRa$s;  dust would  obscure the  star  at intervals
approximately equal  to the pulsation period.  In  the transition from
O--rich to C--rich  stars the amount of dust  reduces, therefore these
objects would behave like  $SRb$s.  Another possibility is that $SRb$s
are less evolved than 0--rich $SRa$s and have therefore have less 
matter  that  can condense  into  dust,  main  requirement to  have  a
multi--periodic light--curve  such as  those calculated by  Winters et
al.~(\cite{wint}).  The monitoring  of  dust features  for $SRa$s  and
$SRb$s  of  comparable near--infrared  properties  will certainly  put
constraints  on  the  proposed  scenarios and  allow  to  discriminate
against the existence of a  binary system; such a program is currently
under development.

Comparing   the  $Log(P)$--magnitude  diagram   with  the   models  by
Vassiliadis \&  Wood (\cite{vass}) we find  that  our sample contains
AGB stars  with masses in the  range from $0.9$ to  $5$ $M_{\sun}$ and
age range  from $0.1$ to  $10$ Gyr. An intermediate/old  population is
definitely present in the LMC.

It is interesting to notice that there are some Semi--Regulars and few
Miras located below  the TRGB. These sources could  be AGB stars
in the  early evolutionary  phase (burning He  in a shell)  or 
TP--AGB (burning H and He into concentric shells) between thermal pulses   
or variables for the
first  time on  the red  giant branch.   We are  going  to investigate
spectroscopically the nature of these variable objects.

\section{Summary}
Based on the cross--identification between the DCMC and the EROS data,
in a  region containing the OC  field in the  bar of the LMC,  we have
classified the best detected variables and shown three $Log(P)$--$K_S$
relation for red giant variables. 
Combined   spectroscopic  observations  have   shown  the
distinction   between   C--rich   and   O-rich  AGB   stars   in   the
color--magnitude  diagram ($J-K_S$, $K_S$).  C--rich stars  occupy the
upper position  of the  $Log(P)$--$K_S$ relations.  We  have confirmed
three of the  relations found by Wood using MACHO  data. The nature of
the  relation at  the  longest  period is  still  unknown and  further
observational investigations are necessary and under development.

\begin{acknowledgements}
The access  to the EROS database  has been kindly granted  by the EROS
collaboration. The GAIA and  CURSA packages have been available thanks
to  the work  of the  Starlink project,  UK.  We  thank Jean--Philippe
Beaulieu,   Sylvain  Lupone  and   Gregoor  Verschuren   for  valuable
assistance  during data  treatment,  Ariane Lan\c  con, Yvonne  Simis,
Peter Wood, Jan--Martin Winters Joris Blommaert and Patricia Whitelock
for useful discussions.
\end{acknowledgements}

\appendix
\section{Tables}

\begin{table}
	\caption{}
	\label{class}
	\[
	\begin{array}{rl}
	\hline
	\noalign{\smallskip}
	\mathrm{Value}  & \mathrm{Description} \\
        \noalign{\smallskip}
        \hline
        \noalign{\smallskip}
 0\,\,\, & \mathrm{Cepheid}\\
11\,\,\, & \mathrm{Mira}\\
13\,\,\, & \mathrm{Mira\,with\,bump}\\
15\,\,\, & \mathrm{SRa}\\
16\,\,\, & \mathrm{SRa\,with\,bump}\\
19\,\,\, & \mathrm{SR}\\
 2\,\,\, & \mathrm{SRb}\\
	\noalign{\smallskip}
        \hline
        \end{array}
        \]
\end{table}

The table (A.2) listing the  DENIS/EROS sources (Fig.~\ref{per}) with a good
light--curve  quality, classified  in the  present work  with observed
spectra is electronically available.  It contains the following items:
the DCMC  source identifier, the  EROS source number, the  period, the
light--curve quality factor, the  classification as described in table
\ref{class}, the EROS quadrant where the source was observed and notes
that could be the Blanco et al. (\cite{bmb}) counterpart or the longer
period in the case of $SRb$s..

Also, a table (A.3) listing the  spectral nature of the newly observed DENIS
 sources, not included in Fig.~\ref{per} and in the previous table, is
 available at the same place. A more detailed analysis with the aim to
 assign the  spectral type to these  sources will be the  subject of a
 forthcoming  paper, together  with equivalent  observations  in other
 locations in the LMC and in the Small Magellanic Cloud.

\end{document}